\documentclass[prb,twocolumn,twocolumngrid,floatfix]{revtex4}
\newcommand{\witheps}[1]{}
\newcommand{\withpdf}[1]{#1}

\usepackage{graphicx}
\usepackage{amsfonts}
\usepackage{amsmath}
\usepackage{bm}
\usepackage{alltt}
\usepackage{fancyhdr}

\begin{document}

\title{Searching for low-workfunction phases in the Cs-Te system: \\ 
the case of Cs$_{2}$Te$_{5}$ }

\author{Anthony Ruth$^{1}$, K\'aroly N\'emeth$^{1,2,{\ast}}$, Katherine C. Harkay$^{2}$, 
Joseph Z. Terdik$^{2}$, Jeff Terry$^{1}$, Linda Spentzouris$^{1,2}$}

\affiliation{$^1$Physics Department, Illinois Institute of Technology, Chicago, IL 60616 USA}

\affiliation{$^2$Advanced Photon Source, Argonne National Laboratory,
Argonne, Illinois 60439, USA}

\date{\today}

\begin{abstract}
We have computationally explored workfunction values of Cs$_{2}$Te$_{5}$, an existing crystalline
phase of the Cs-Te system and a small bandgap semiconductor, in order to search for
reduced workfunction alternatives of Cs$_{2}$Te  
that preserve the exceptionally high quantum efficiency of the Cs$_{2}$Te seasoned photoemissive
material. We have found that the Cs$_{2}$Te$_{5}$(010) surface exhibits a workfunction value of 
$\approx$ 1.9 eV when it is covered by Cs atoms. 
Cs$_{2}$Te$_{5}$ is analogous to our recently proposed
low-workfunction materials, Cs$_{2}$TeC$_{2}$ and other
ternary acetylides [J. Z. Terdik, {\it et al.}, Phys. Rev. B 86, 035142 (2012)], 
in as much as it also contains quasi one-dimensional substructures
embedded in a Cs-matrix, forming the foundation for anomalous workfunction anisotropy, and low
workfunction values. The one-dimensional substructures in Cs$_{2}$Te$_{5}$ are polytelluride ions
in a tetragonal rod packing. Cs$_{2}$Te$_{5}$ has the advantage of simpler composition and
availability as compared to Cs$_{2}$TeC$_{2}$, however its low workfunction surface is less
energetically favored to the other surfaces than in Cs$_{2}$TeC$_{2}$.
\end{abstract}


\maketitle
\section{Introduction}
Cesium Telluride (Cs$_{2}$Te) has been known since the 1950-s as an exceptionally high quantum
efficiency photoemissive material \cite{ETaft53}, it can turn as much as $\approx$20\% of the incident 
ultraviolet photons into emitted electrons \cite{DVelazquez12}. Cesium Telluride also has the advantage
of relatively long operational lifetime, 20-30 times longer than that of competing multi-alkali
antimonide photocathodes, such as K$_{2}$CsSb and (Cs)Na$_{3}$KSb. While K$_{2}$CsSb and (Cs)Na$_{3}$KSb
require ultrahigh vacuum for operation, Cs$_{2}$Te can operate in orders of magnitudes lower levels of
vacuum \cite{DHDowell10}. 
In order to further enhance the photoemissive properties of Cs$_{2}$Te for certain applications, 
modifications are required that
decrease its workfunction from the $\approx$3.0 eV down to the visible light spectrum (1.5-3.0 eV) while
preserving its high quantum efficiency. Such modifications can lead to, for example, for high brightness
electron guns \cite{DHDowell10,KNemeth10}, better pulse shaping
of the incident photons in the visible spectrum and eliminating the need for wavelength down-conversion.
Wavelength down-conversion is used to convert the typically near-infrared
photons of the laser source to ultraviolet wavelength which causes a great 
loss of the intensity of the initial laser-beam. One possible way to an improved
photoemissive material that we recently proposed \cite{JZTerdik12}
is the acetylation of Cs$_{2}$Te leading to Cs$_{2}$TeC$_{2}$, a new member of the existing family of
ternary acetylide \cite{URuschewitz06,URuschewitz01A,HBilletter10} compounds. Electronic structure
calculations predict that the new Cs$_{2}$TeC$_{2}$ and other, existing ternary acetylides, such as
Cs$_{2}$PdC$_{2}$ would have similarly high quantum efficiencies as Cs$_{2}$Te, 
but significantly lower, 2.0-2.4 eV workfunctions.

\begin{figure}[b!]
\withpdf{\resizebox*{3.4in}{!}{\includegraphics{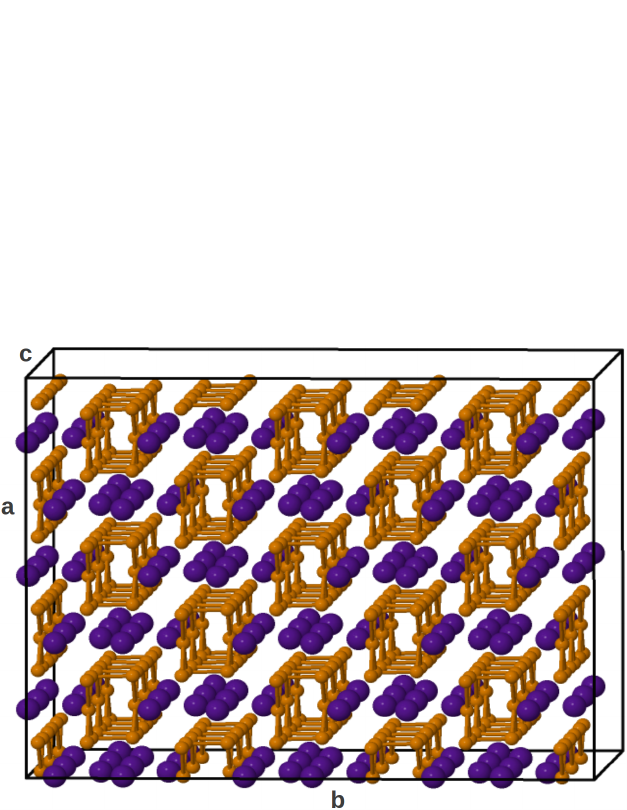}}}
\witheps{\resizebox*{3.4in}{!}{\includegraphics{Fig1.eps}}}
\caption{
A near top down view of a 3x3x3 supercell of the rectangular unit cell of Cs$_{2}$Te$_{5}$.
Cell data are from Ref. \onlinecite{PBoettcher82}.
Bronze spheres denote Te, blue ones are Cs. Notice the quasi 1D [Te$_{5}^{2-}$]$_{n}$ polytelluride
ions embedded in Cs matrix in the form of $\approx$ 4 {\AA} wide wavy Te-ribbons,
showed in detail in Fig. \ref{Te-ribbon}.
}
\label{Cs2TeC5unitcell}
\end{figure}

\begin{figure}[t!]
\withpdf{\resizebox*{3.4in}{!}{\includegraphics{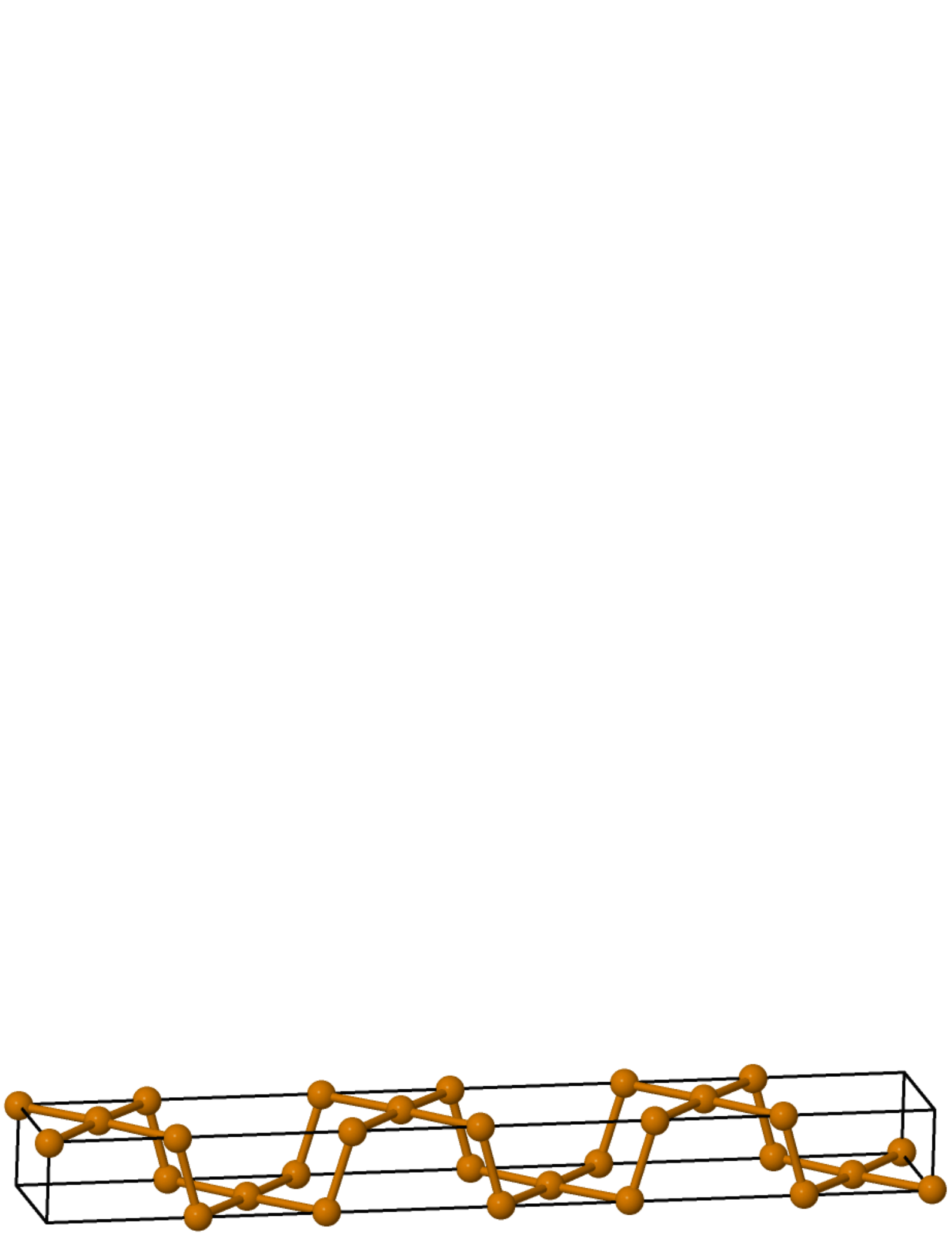}}}
\witheps{\resizebox*{3.4in}{!}{\includegraphics{Fig2.eps}}}
\caption{
An isolated Te-ribbon of the 3x3x3 supercell of Cs$_{2}$Te$_{5}$.
In the wavy Te-ribbons, six-membered rings of Te in
chair-conformation are connected via common vortices into quasi 1D chains.
}
\label{Te-ribbon}
\end{figure}

An alternative route to the acetylation in developing improved photoemissive analogues/derivatives 
of Cs$_{2}$Te might be in the exploration of photoemissive properties of other Cs-Te phases.
A comprehensive review of alkali tellurides by D. M. Smith and J. A. Ibers \cite{DMSmith00} called our
attention to Cs$_{2}$Te$_{5}$, an existing \cite{PBoettcher82} 
crystalline phase of Cs and Te. Remarkably, the
Te$_{5}^{2-}$ polytelluride anions in Cs$_{2}$Te$_{5}$ self organize to 
$\approx$ 4 {\AA} wide wavy ribbons of Te with continuous covalent Te-Te networks, 
which are embedded into a Cs matrix, such as shown in Fig. \ref{Cs2TeC5unitcell}. 
In the wavy Te-ribbons, six-membered rings of Te in
chair-conformation are connected via common vortices into quasi 1D chains,
as depicted in Fig. \ref{Te-ribbon}.
These quasi 1D substructures of Cs$_{2}$Te$_{5}$ resemble the rod-like
polymeric [-Te-C$\equiv$C-]$_{n}^{2n-}$ substructures that are responsible for the improved photoemissive
properties of Cs$_{2}$TeC$_{2}$. This structural analogy directed our attention towards the computational 
analysis of Cs$_{2}$Te$_{5}$ to check whether it can potentially serve as 
an improvement to Cs$_{2}$Te and ternary acetylides. 

%
\begin{figure}[tb!]
\withpdf{\resizebox*{3.4in}{!}{\includegraphics{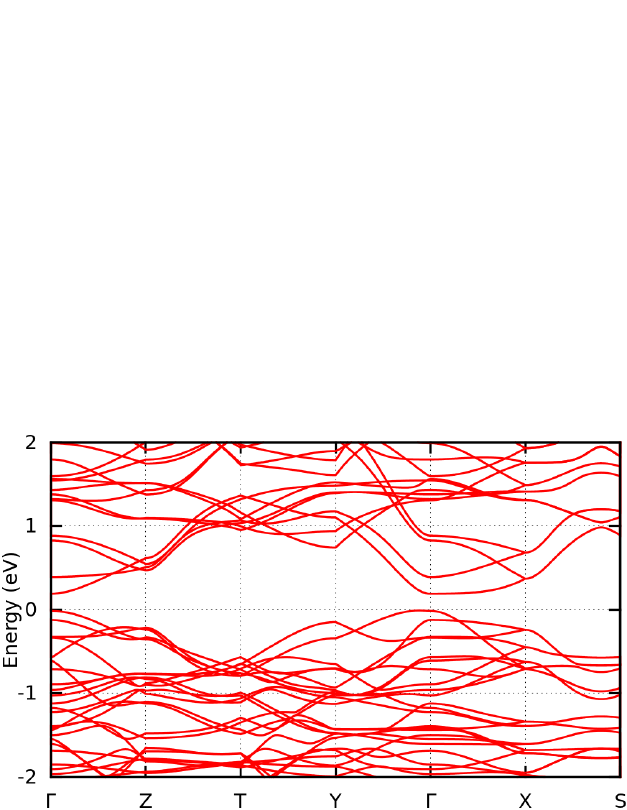}}}
\witheps{\resizebox*{3.4in}{!}{\includegraphics{Fig3.eps}}}
\caption{
Band structure of Cs$_{2}$Te$_{5}$. Energy levels are relative to the top of the valence
band. The band gap is 0.19 eV. 
The selection of the special k-points is based on the orthorombic symmetry of the cell and is identical
with that used for Cs$_{2}$Te in Ref. \onlinecite{JZTerdik12}, as both Cs$_{2}$Te (space group: Pnma) and 
Cs$_{2}$Te$_{5}$ (space group: Cmcm) crystallize in the orthorombic system.
}
\label{Cs2Te5bands}
\end{figure}

\section{Methodology}

The electronic structure calculations in the present study have been carried out using the 
Quantum Espresso program package \cite{QE}. 
The PBE exchange-correlation potential \cite{PBE} has been used 
with norm-conserving Cs and Te pseudopotentials identical
to those in our study for Cs$_{2}$TeC$_{2}$ in Ref. \onlinecite{JZTerdik12}.
The wavefunction-cutoff was $80$ Rydbergs. 
The k-space grids were at least 6$\times$6$\times$6 large for optimizations, the residual forces on
fractional
coordinates were less than $4{\times}10^{-4}$ Ry/au, residual pressure on the unit cell
less than 1 kbar.
The application of these pseudopotentials has been  
validated on known properties (structure and workfunction) of crystalline Cs, Te and Cs$_{2}$Te.
This method provides a=9.719, b=12.178 and c=10.407 {\AA} 
for the calculated lattice parameters of Cs$_{2}$Te$_{5}$, while the
experimental ones \cite{PBoettcher82} are a=9.373, b=12.288 and c=10.140 {\AA}, an agreement within
3.5\% . There are two different Te-Te bond lengths in Cs$_{2}$Te$_{5}$: the calculated values
are 2.845 and 3.090 {\AA}, the experimental ones are 2.765 and 3.049 {\AA}, respectively, an agreement
within 3\%. Experimental and calculated lattice angles are all 90$^{o}$.
Optical absorption spectra have been calculated in the Random Phase Approximation (RPA) as
implemented in the YAMBO \cite{AMarini09} code, the spectra were calculated using a gradually
increased number of planewaves and interaction block-size until convergence at 20000 plane waves and
an interaction block size of 403. 
For comparison, data for Cs$_{2}$Te and Cs$_{2}$TeC$_{2}$ have been taken from Ref. \onlinecite{JZTerdik12}.
The workfunction calculations were based on slabs of at least 30 {\AA} width separated by
vacuum layers up to 30 {\AA} following the methodology of Ref. \onlinecite{CJFall99}.
Only the top and bottom two layers were relaxed while middle layers were kept at the bulk optimum
structures.

The two most important crystal surfaces of Cs$_{2}$Te$_{5}$ that do not cleave the polytelluride
anions and have Miller indices of (010) and (110) have been considered. 
The cleavage of polytelluride anions is energetically unfavored and would lead to a plethora of
possible surface reconstructions. To estimate the strength of the simplest such cleavage, properties
of the (001) surface have also been calculated.
For the (010) surfaces, two different cleavages
have been modeled, (010)-C1 and (010)-C2. The (110) and (010)-C1 surface slabs have identical
top and bottom surfaces, while the (010)-C2 surfaces are not identical. The (010)-C1 surface
slab leaves some Te atoms directly exposed on both of its surfaces, while 
the (010)-C2 one has one fully Cs covered surface and one partially Cs covered one.

\begin{table}[tb!]
\caption{
Calculated properties of Cs$_{2}$Te$_{5}$ surfaces:
workfunctions ($\Phi$), bandgaps at the $\Gamma$-point
E$_{g}(\Gamma)$ and surface energies ($\sigma$).
For the Cs$_{2}$Te$_{5}$(010)-C2 cleavage, data refer to 
the fully Cs-covered surface. The workfunction of this surface
has been calculated both from the asymmetrically Cesiated 
Cs$_{2}$Te$_{5}$(010)-C2 slab ($\Phi=$ 1.87 eV)
and from the symmetrized (with additional Cs) and relaxed version of it 
($\Phi=$ 1.97 eV). The average surface energy of the asymmetrically Cesiated 
Cs$_{2}$Te$_{5}$(010)-C2 slab was 22.6 meV/{\AA}$^{2}$, 
the contribution of the Cs-rich side is estimated to be close to the
Cs$_{2}$Te$_{5}$(010)-C1 value ($\sigma=$ 7.1 meV/{\AA}$^{2}$).
}
\label{Workfunctions}
\begin{tabular}{cccc}
\toprule
         & $\Phi$  & E$_{g}(\Gamma)$  & $\sigma$  \\
surface  & (eV) & (eV) & (meV/{\AA}$^{2}$)  \\
\colrule
Cs$_{2}$Te$_{5}$(110)       & 3.22       & 0.3577 &  7.2    \\
Cs$_{2}$Te$_{5}$(010)-C1    & 3.47       & 0.3344 &  7.1    \\
Cs$_{2}$Te$_{5}$(010)-C2    & 1.87/1.97  & 0.0309 &  -      \\
Cs$_{2}$Te$_{5}$(001)       & 4.70       & 0.0369 & 20.4    \\
\botrule
\end{tabular}
\end{table}

\section{Results and Discussion}
\begin{figure}[t!]
\withpdf{\resizebox*{3.4in}{!}{\includegraphics{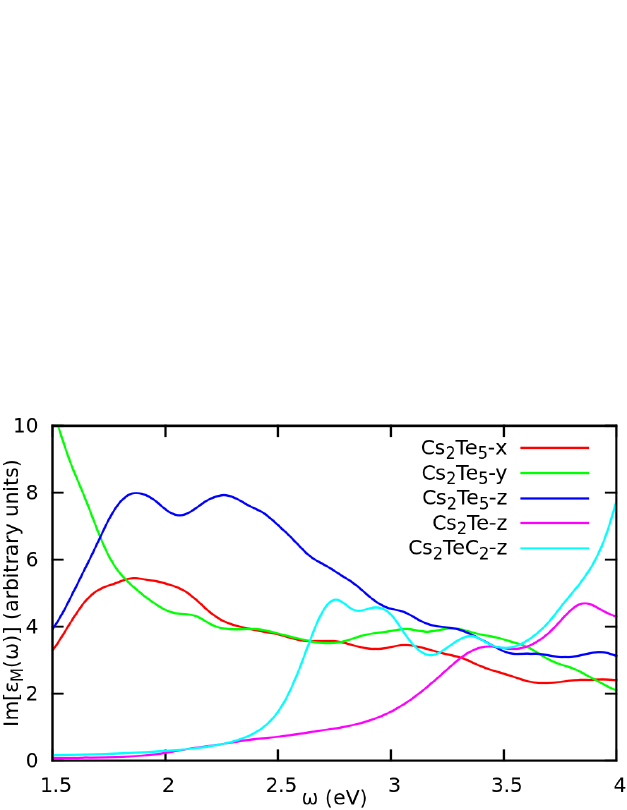}}}
\witheps{\resizebox*{3.4in}{!}{\includegraphics{Fig4.eps}}}
\caption{
Optical absorption spectra in terms of the 
macroscopic dielectric constant $\varepsilon_{M}(\omega)$
for Cs$_{2}$Te$_{5}$ as compared to that of Cs$_{2}$Te and Cs$_{2}$TeC$_{2}$.
The energy of the incident photons is denoted by 
$\omega$, while the polarization of the photons is
indicated by the coordinate directions in the curve-keys with $z$ being parallel with the main
crystallographic axis. 
The calculations predict that Cs$_{2}$Te$_{5}$ would have a significantly higher absorption
probability at lower photon energies, also at $\approx$ 1.9 eV which is the workfunction value of the
Cs-covered Cs$_{2}$Te$_{5}$(010) surface.
}
\label{Cs2Te5optabs}
\end{figure}

The bandstructure of Cs$_{2}$Te$_{5}$ is shown in Fig. \ref{Cs2Te5bands}. It indicates a band gap
of 0.19 eV, i.e. a small bandgap semiconductor material. 
The optical absorption spectra of Cs$_{2}$Te$_{5}$ with different light polarizations are presented
in Fig. \ref{Cs2Te5optabs} and are compared to those of Cs$_{2}$Te and Cs$_{2}$TeC$_{2}$.
The small gap value and the strong absorption at low photon energies appears to be consistent
with the experimentally observed ``metallic grey'' color \cite{PBoettcher82} of Cs$_{2}$Te$_{5}$.
The workfunction values for several important crystallographic 
surfaces of Cs$_{2}$Te$_{5}$ are listed in Table \ref{Workfunctions}.
It is apparent that the lowest workfunction belongs to the Cs$_{2}$Te$_{5}$(010)-C2 
surface with fully Cs-covered polytelluride ribbons. The other surfaces have 
much greater workfunctions, in the 3.2-3.5 eV range or even above for the Cs$_{2}$Te$_{5}$(001)
cleavage. Also note that cleaving the polytelluride ribbons results in a relatively large surface energy,
i.e. it is energetically very unfavorable as compared to the other cleavages.

%


Comparing the workfunction value of $\approx$ 1.90 eV for the Cs-rich one of the Cs$_{2}$Te$_{5}$(010)-C2 surfaces 
to the optical absorption spectra in Fig. \ref{Cs2Te5optabs} one can see a very strong absorption at 1.9 eV.
This means that if this surface can be realized, it will have not only a significantly (by more than 1 eV) 
reduced workfunction, but also an even higher optical absorption than Cs$_{2}$Te, and as a consequence its
quantum yield may be higher than that of Cs$_{2}$Te. At the same time, Cs$_{2}$Te$_{5}$ is just a
two-component material and its synthesis is readily available from the literature \cite{PBoettcher82}.
It is however also to be noted that the surface energy differences between the (010) and (110) surfaces are
very small, making it difficult to realize the Cs-covered Cs$_{2}$Te$_{5}$(010)-C2 cleavage without coexisting
presence of other surfaces. Since large single crystals can be grown of Cs$_{2}$Te$_{5}$ as reported in Ref.
\onlinecite{PBoettcher82}, the suitable cleavage of them could deliver the required surface. 
Since the Cs$_{2}$Te$_{5}$ material is relatively soft due to the weak interaction
between Te-ribbons and the small cleavage energy between Cs-covered Te-ribbons, it
may also be deposited along these easy-to-cleave surfaces by 
rubbing larger crystals of Cs$_{2}$Te$_{5}$ to the
substrate surface, similarly to deposition of graphite.

Bandgaps of the surface slabs at the $\Gamma$-point differ somewhat from the bulk-value (0.19 eV) in the
range of $\approx$ 0.037-0.360 eV, being smallest for the lowest workfunction Cs$_{2}$Te$_{5}$(010)-C2 slab. 

While the acetylated Cs$_{2}$Te, i.e. Cs$_{2}$TeC$_{2}$, shows very anisotropic workfunction values,
surface energies and optical absorptions (a factor of 9 optical absorption differences \cite{JZTerdik12}), 
the degree of anisotropy is somewhat smaller for
Cs$_{2}$Te$_{5}$, where optical absorptions may differ by a factor of two in the visible spectral range.

\begin{figure}[t!]
\withpdf{\resizebox*{3.4in}{!}{\includegraphics{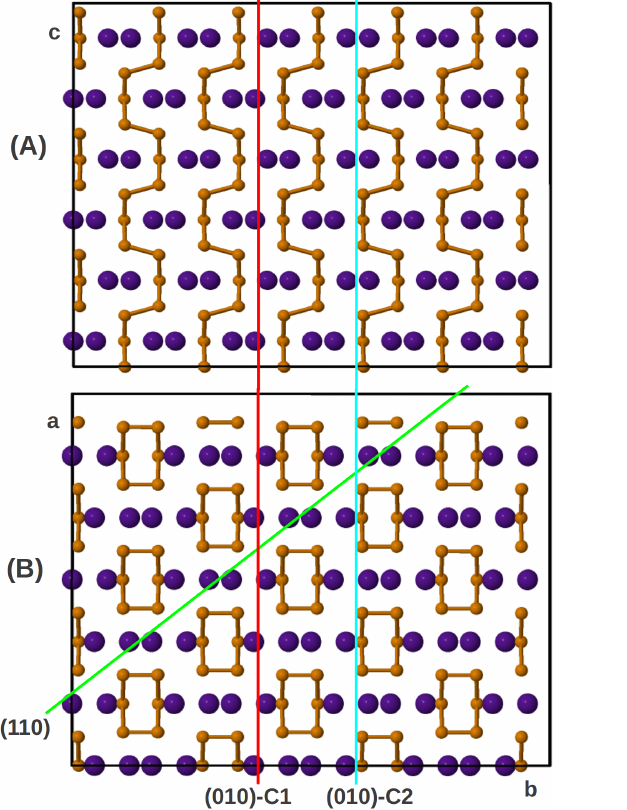}}}
\witheps{\resizebox*{3.4in}{!}{\includegraphics{Fig5.eps}}}
\caption{
Panels (A) and (B) show the (bc) and (ba) plane views of a 3x3x3 supercell of the
Cs$_{2}$Te$_{5}$ crystal, respectively, where a, b and c denote the crystallographic axes.
Only those crystal surfaces have been considered that do not cleave polytelluride ions and have
small Miller indices. These are (110), (010)-C1 and (010)-C2. 
The (010)-C1 surface slab leaves some Te atoms directly exposed on both of its surfaces, while 
the (010)-C2 one has one fully Cs covered surface and one partially Cs covered one.
}
\label{Cs2Te5surfaces}
\end{figure}
%

\section{Conclusions}
In the present study we have computationally analyzed bulk and surface properties of Cs$_{2}$Te$_{5}$, 
an existing \cite{PBoettcher82} crystalline phase of the Cs-Te system, in order to search
for alternatives of Cs$_{2}$Te in the Cs-Te system with reduced workfunctions and high quantum
efficiency. We have found that the fully Cs-covered Cs$_{2}$Te$_{5}$(010)-C2 surface
has a workfunction of $\approx$ 1.9 eV and a quantum efficiency that is higher than that of Cs$_{2}$Te, both 
at the respective workfunction values. Since large single crystals of Cs$_{2}$Te$_{5}$ can be produced
as described in Ref. \onlinecite{PBoettcher82}, this prediction can be validated experimentally, and
Cs$_{2}$Te$_{5}$ can become a practical alternative to Cs$_{2}$Te for photophysical applications.
 
\section{Acknowledgements}
The authors gratefully acknowledge A. Zholents, Z. Yusof and K. Attenkofer (APS/Argonne)
for helpful discussions and thank NERSC (U.S. DOE DE-AC02-05CH11231) for
the use of computational resources. 
This research was supported
by the U.S. DOE Office of Science, under contract No.
DE-AC02-06CH11357, and also by the National Science Foundation (No. PHY-0969989).

\noindent
{$\ast$ Nemeth@ANL.Gov}
%
%

\end{document}